%% file: corfu_arXiv.tex
\documentclass{PoS_arXiv}

\usepackage{amsmath,amssymb,amsfonts,color,graphicx,cite,color,soul}
\usepackage{pifont}
\newcommand{\cmark}{\ding{51}}%
\newcommand{\xmark}{\ding{55}}%
\usepackage{float}
\usepackage[numbers,sort&compress]{natbib}
\usepackage{url}
\title{An N2HDM Solution for the possible 96 GeV Excess}

\ShortTitle{An N2HDM Solution for the possible 96 GeV Excess}

\author{T.~Biek\"otter\\
        Instituto de F\'isica Te\'orica, (UAM/CSIC) \& \\
        Departamento de F{\'i}sica Te{\'o}rica, Universidad
        Aut\'onoma de Madrid,\\  
        Cantoblanco, E-28049 Madrid, Spain\\
        E-mail: \email{thomas.biekotter@csic.es}}

\author{M.~Chakraborti\\
        Instituto de F\'isica Te\'orica, (UAM/CSIC) 
        Universidad Aut\'onoma de Madrid,\\  
        Cantoblanco, E-28049 Madrid, Spain\\
        E-mail: \email{mani.chakraborti@gmail.com}}

\author{\speaker{S.~Heinemeyer}\\ 
        Instituto de F\'isica Te\'orica, (UAM/CSIC), Universidad
  Aut\'onoma de Madrid,\\  
Cantoblanco, E-28049 Madrid, Spain\\
Campus of International Excellence UAM+CSIC, Cantoblanco, E-28049,
Madrid, Spain\\
Instituto de F\'isica de Cantabria (CSIC-UC), E-39005 Santander, Spain\\
        E-mail: \email{Sven.Heinemeyer@cern.ch}}

\abstract{We discuss a $\sim 3\,\sigma$
signal (local) in the light Higgs-boson search
in the diphoton decay mode at $\sim 96 \gev$ as reported by
CMS, together with a $\sim 2\,\sigma$ excess (local)
in the $b \bar b$ final state
at LEP in the same mass range.
We review the interpretation of this possible signal as a Higgs boson in the
2~Higgs Doublet Model with an additional real Higgs singlet (N2HDM).
It is shown that the lightest Higgs boson of the N2HDM can perfectly fit
both excesses simultaneously, while the full Higgs-boson
sector is in agreement with all Higgs-boson measurements and exclusion
bounds as well as other theoretical and
experimental constraints. It is demonstrated that in particular the N2HDM
type~II and can fit the data best, leading to a supersymmetric
interpretation. The NMSSM and the \mnSSM\ are briefly reviewed in this
respect. 
}

\FullConference{
Corfu Summer Institute 2018 "School and Workshops on Elementary Particle
Physics and Gravity"\\ 
                (CORFU2018)\\
                31 August - 28 September, 2018\\
                Corfu, Greece}

\include{paperdef}

\graphicspath{{figs/}}

\begin{document}


\section {Introduction}
\label{sec:intro}

The Higgs boson discovered in 2012 by ATLAS and
CMS~\cite{Aad:2012tfa,Chatrchyan:2012xdj} -- within theoretical and
experimental uncertainties -- is consistent with the existence of a
Standard-Model~(SM) Higgs boson~\cite{Khachatryan:2016vau}.
However, the measurements of Higgs-boson couplings, which are known
experimentally to a precision of roughly $\sim 20\%$, leave room for
Beyond Standard-Model (BSM) interpretations. Many BSM models possess
extended Higgs-boson sectors. which naturally contain additional Higgs
bosons with 
masses larger than $125 \gev$. However, many extensions also offer the
possibilty of additional Higgs bosons {\em below} 
$125 \gev$. Consequently, the search for lighter Higgs bosons forms an
important part in the BSM Higgs-boson analyses.

Searches for Higgs bosons below $125 \gev$ have been performed at LEP,
the Tevatron and the LHC. 
LEP reported a $2.3\,\sigma$ local excess
observed in the~$e^+e^-\to Z(H\to b\bar{b})$
searches\,\cite{Barate:2003sz}, which would be consistent with a
scalar of mass ~$\sim 98 \gev$, but due to the $b \bar b$ final state the 
mass resolution is rather coarse). The excess corresponds to 
\begin{equation}
\mu_{\rm LEP}=\frac{\sigma\left( e^+e^- \to Z \phi \to Zb\bar{b} \right)}
			   {\sigma^{\mathrm{SM}}\left( e^+e^- \to Z H
			   		\to Zb\bar{b} \right)}
			  = 0.117 \pm 0.057 \; ,
\label{muLEP}
\end{equation}
where the signal strength $\mu_{\rm LEP}$ is the
measured cross section normalized to the SM expectation,
with the SM Higgs-boson mass at $\sim 98\gev$.
The value for $\mu_{\rm LEP}$ was extracted in \citere{Cao:2016uwt}
using methods described in \citere{Azatov:2012bz}.

Recent CMS~Run\,II
results\,\cite{Sirunyan:2018aui} for Higgs-boson searches in the diphoton
final state show a local excess of~$\sim 3\,\sigma$ around
$\sim 96 \gev$, with a similar excess
of~$2\,\sigma$ in the Run\,I data at a comparable
mass~\cite{CMS:2015ocq}.
The excess corresponds to (combining 7, 8 and $13 \tev$
data, and assuming that the $gg$ production dominates)
\begin{equation}
\mu_{\rm CMS}=\frac{\sigma\left( gg \to \phi \to \gamma\gamma \right)}
         {\sigma^{\rm SM}\left( gg \to H \to \gamma\gamma \right)}
     = 0.6 \pm 0.2 \; .
\label{muCMS}
\end{equation}
First Run\,II~results from~ATLAS
with~$80$\,fb$^{-1}$ in the~$\ga\ga$~searches below~$125$\,GeV turned
out to be weaker than the corresponding CMS results, see, e.g., Fig.~1
in \citere{Heinemeyer:2018wzl}. 

Reviews about the possibility that these two
excesses, found effectively at the same mass, are of a common origin.
are given in \citeres{Heinemeyer:2018jcd,Heinemeyer:2018wzl}. The list
comprises of type~I 2HDMs~\cite{Fox:2017uwr,Haisch:2017gql}, a radion
model~\cite{Richard:2017kot},
a minimal dilaton model~\cite{LiuLiJia:2019kye},
as well as supersymmetric
models~\cite{Biekotter:2017xmf,Domingo:2018uim,Hollik:2018yek}.

\medskip
Motivated by the Hierarchy Problem, Supersymmetry (SUSY) plays a
prominent role in BSM physics. The simplest SUSY extension of the SM is
the Minimal Supersymmetric Standard Model
(MSSM)~\cite{Nilles:1983ge,Haber:1984rc}, doubling the degrees of freedom
of the SM supplemented with a second Higgs doublet. 
The MSSM Higgs sector, composed of $\Phi_1$ and $\Phi_2$, 
consists of two $\cp$-even, one
$\cp$-odd and two charged Higgs bosons. The light (or the heavy)
$\cp$-even MSSM Higgs boson can be interpreted as the signal discovered
at $\sim 125 \gev$~\cite{Heinemeyer:2011aa} (see
\citeres{Bechtle:2016kui,Bahl:2018zmf} for recent updates). However,
in \citere{Bechtle:2016kui} it was demonstrated that the MSSM cannot
explain the CMS excess in the diphoton final state. This can be traced
back to the ``too rigid'' structure of the 2HDM (type~II) strucure of the
Higgs-boson sector in the MSSM. 

This raises the question whether simple extensions of the 2HDM can fit
both the CMS excess in \refeq{muCMS} and the LEP exceses in \refeq{muLEP}.
In \citere{Biekotter:2019kde} the Next to minimal 2 Higgs doublet model
(N2HDM)~\cite{Chen:2013jvg,Muhlleitner:2016mzt} was investigated. 
In this model the two Higgs doublets are
supplemented with a real Higgs singlet, giving rise 
to one additional (potentially light) $\cp$-even Higgs boson. However, 
in comparison to SUSY models the N2HDM does not have to
obey the SUSY relations in the Higgs-boson sector. Consequently, it
allows to study how the potential fits the two excesses
simultaneously in a more general way. Here we review first the results
obtained in the N2HDM~\cite{Biekotter:2019kde} and then two possible
SUSY realizations.


\section{The N2HDM, constraints and the experimental excesses}
\label{sec:model}

\subsection{The N2HDM}

The N2HDM is the simplest 
extension of a $\cp$-conserving two Higgs doublet model (2HDM) 
in which the latter is augmented with a real 
scalar singlet Higgs field, denoted as 
$\Phi_1$, $\Phi_2$ and $\Phi_S$, respectively
(see, e.g., \citeres{Chen:2013jvg,Muhlleitner:2016mzt}).
As in the 2HDM a $Z_2$ symmetry is imposed to avoid flavor changing
neutral currents at the tree-level,
only softly broken in the Higgs sector via the bilinear
mass term $m_{12}^2(\Phi_1^\dagger \Phi_2 + h.c.)$. 
As in the 2HDM, this leads to four variants of 
the N2HDM, depending on the $Z_2$ parities of the 
fermions. Taking the electroweak symmetry breaking (EWSB) minima to be
charge and $\cp$-conserving, the scalar fields after EWSB
can be parametrised as
\begin{eqnarray}
\Phi_1 = \left( \begin{array}{c} \phi_1^+ \\ \frac{1}{\sqrt{2}} (v_1 +
    \rho_1 + i \eta_1) \end{array} \right) \;, \quad
\Phi_2 = \left( \begin{array}{c} \phi_2^+ \\ \frac{1}{\sqrt{2}} (v_2 +
    \rho_2 + i \eta_2) \end{array} \right) \;, \quad
\Phi_S = v_S + \rho_S \;, \label{eq:n2hdmvevs}
\end{eqnarray}
where $v_1, v_2, v_S$ are the real vevs acquired by the fields
$\Phi_1, \Phi_2$ and $\Phi_S$ respectively.
As in the 2HDM we define $\tb := v_2/v_1$.
The $\cp$-even Higgs-boson sector contains three physical Higgses. Thus,
a rotation from the 
interaction to the physical basis can be achieved with the help of
a $3 \times 3$ orthogonal matrix as

\begin{eqnarray}
\left( \begin{array}{c} h_1 \\ h_2 \\ h_3 \end{array} \right) = R
\left( \begin{array}{c} \rho_1 \\ \rho_2 \\ \rho_S \end{array} \right)~,
\end{eqnarray}
with $m_{h_1} < m_{h_2} < m_{h_3}$. 
The rotation matrix $R$ can be parametrized as
\begin{equation}
\label{mixingmatrix}
R=
\begin{pmatrix}
c_{\alpha_1}c_{\alpha_2} &
  s_{\alpha_1}c_{\alpha_2} &
    s_{\alpha_2} \\
-(c_{\alpha_1}s_{\alpha_2}s_{\alpha_3}+s_{\alpha_1}c_{\alpha_3}) &
  c_{\alpha_1}c_{\alpha_3}-s_{\alpha_1}s_{\alpha_2}s_{\alpha_3}  &
    c_{\alpha_2}s_{\alpha_3} \\
-c_{\alpha_1}s_{\alpha_2}c_{\alpha_3}+s_{\alpha_1}s_{\alpha_3} &
-(c_{\alpha_1}s_{\alpha_3}+s_{\alpha_1}s_{\alpha_2}c_{\alpha_3}) &
c_{\alpha_2}c_{\alpha_3}
\end{pmatrix}~,
\end{equation}
$\alpha_1, \alpha_2, \alpha_3$ being the three mixing angles, and
we use the short-hand notation $s_x = \sin x$, $c_x = \cos x$.
The couplings of the Higgs bosons to SM particles are modified
w.r.t.\ the SM Higgs-coupling predictions due to the mixing in the Higgs
sector. It is convenient to express the couplings of the scalar
mass eigenstates $h_i$ normalized to the corresponding SM couplings.
We therefore introduce the coupling coefficients $c_{h_i V V}$ and
$c_{h_i f \bar f}$, such that the couplings to the massive vector bosons
are given by
\begin{equation}
\left(g_{h_i W W}\right)_{\mu\nu} =
\mathrm{i} g_{\mu\nu} \left(c_{h_i V V}\right) g M_W
\quad \text{and } \quad
\left(g_{h_i Z Z}\right)_{\mu\nu} =
\mathrm{i} g_{\mu\nu} \left(c_{h_i V V}\right) \frac{g M_Z}{\CW} \, ,
\end{equation}
where $g$ is the $SU(2)_L$ gauge coupling, $\CW$ the cosine of weak
mixing angle, $\CW = \MW/\MZ, \SW = \sqrt{1 - \CW^2}$,
and $M_W$ and $M_Z$ the masses of the $W$ boson
and the $Z$ boson, respectively. The couplings of the Higgs bosons
to the SM fermions are given by
\begin{equation}
g_{h_i f \bar{f}} =
\frac{m_f}{v} \left(c_{h_i f \bar{f}}\right) \; ,
\end{equation}
where $m_f$ is the mass of the fermion and
$v = \sqrt{(v_1^2 +v_2^2)}$ is the SM vev.
The coupling coefficients for the couplings to
gauge bosons $V = W,Z$ for the three $\cp$-even Higgses.
are identical in all four types of the (N)2HDM.
They differ, however, as in the 2HDM depending on the type of the model,
as summarized in \refta{tab:hff}.

\begin{table}
\centering
\begin{tabular}{lccc} \hline
& $u$-type ($c_{h_i t\bar t}$)& $d$-type ($c_{h_i b\bar b}$)&
  leptons ($c_{h_i \tau\bar\tau}$)\\ \hline
type~I & $R_{i2} / s_\beta$
& $R_{i2} / s_\beta$ &
$R_{i2} / s_\beta$ \\
type~II & $R_{i2} / s_\beta $
& $R_{i1} / c_\beta $ &
$R_{i1} / c_\beta $ \\
type~III (lepton-specific) & $R_{i2} / s_\beta$
& $R_{i2} / s_\beta$ &
$R_{i1} / c_\beta$ \\
type~IV (flipped) & $R_{i2} / s_\beta$
& $R_{i1} / c_\beta$ &
$R_{i2} / s_\beta$ \\ \hline
\end{tabular}
\caption{Coupling factors of the Yukawa couplings of
   the N2HDM Higgs bosons $h_i$ w.r.t.\ their SM values.}
\label{tab:hff}
\end{table}

There are 12 independent parameters in the model,
which can be taken as~\cite{Muhlleitner:2016mzt};
\begin{equation}
\alpha_{1,2,3} \; , \quad \tan\beta \;, \quad v \; ,
\quad v_S \; , \quad m_{h_{1,2,3}} \;, \quad m_A \;, \quad \MHp
\;, \quad m_{12}^2 \; , \label{eq:inputs}
\end{equation}
where
$m_A$, $\MHp$ denote
the masses of the physical $\cp$-odd and charged Higgses respectively.

In \citere{Biekotter:2019kde} the code 
\texttt{ScannerS}~\cite{Coimbra:2013qq,Muhlleitner:2016mzt} has been
used to uniformly explore the set of independent parameters as
given in \refeq{eq:inputs} (see below).
The lightest $\cp$-even Higgs boson, $h_1$, was identified with the one
being potentially responsible for the signal at 
$\sim 96 \gev$. The second lightest $\cp$-even Higgs boson was 
identified with the one observed at $\sim 125 \gev$.


\subsection{Constraints}

All relevant constraints on the N2HDM were taken into
account, see \citere{Biekotter:2019kde} for more details. These comprise

\begin{itemize}

\item Theoretical constraints:\\
tree-level perturbativity and the
condition that the vacuum should be a global minimum of
the potential.

\item Constraints from direct searches at colliders:\\
All relevant searches for BSM Higgs bosons are taken into account with
the code \texttt{HiggsBounds v.5.3.2}~\cite{Bechtle:2008jh,Bechtle:2011sb,Bechtle:2013gu,Bechtle:2013wla,Bechtle:2015pma}.

\item Constraints from the SM-like Higgs-boson properties:\\
Any model beyond the SM has to accommodate the SM-like Higgs boson,
with mass and signal strengths as they were measured at the LHC.
In our scans the compatibility of the $\cp$-even scalar $h_2$ with a mass
of $125.09\gev$ with the measurements of signal strengths at Tevatron and LHC
is checked with the code \texttt{HiggsSignals v.2.2.3}~\cite{Bechtle:2013xfa,Stal:2013hwa,Bechtle:2014ewa}.
The corresponding theory predictions are proved by a combination of the
codes \texttt{ScannerS},
\texttt{SusHi}~\cite{Harlander:2012pb,Harlander:2016hcx} and 
\texttt{N2HDECAY}~\cite{Muhlleitner:2016mzt,Djouadi:1997yw,Butterworth:2010ym}.
The \texttt{HiggsSignals} output shown below consists in the reduced $\chi^2$,
\begin{equation}
  \label{eq:chiHS}
  \chi_{\rm red}^2 = \frac{\chi^2}{n_{\rm obs}} \; ,
\end{equation}
where $\chi^2$ is provided by \texttt{HiggsSignals} and $n_{\rm obs}=101$ is
the number of experimental observations considered.

\item Constraints from flavor physics:\\
In the low $\tb$ region that is of interest (see below)
the constraints which must be taken into account 
are \cite{Arbey:2017gmh}: $\br(B \to X_s \gamma)$, 
constraints on $\Delta M_{B_s}$
from neutral $B-$meson mixing and $\br(B_s \to \mu^+ \mu^-)$. 
Constraints from $\br(B \to X_s \gamma)$ 
excludes $\MHp < 650\gev$ for all values of 
$\tan\beta \gtrsim 1$ in the type~II and~IV 2HDM, while for type~I 
and~III the bounds are more $\tan\beta-$dependent.

\item Constraints from electroweak precision data:\\
Constraints from electroweak precision observables can in a simple
approximation be expressed in terms of the oblique parameters S, T and
U~\cite{Peskin:1990zt,Peskin:1991sw}. 
Deviations to these parameters are significant if new physics
beyond the SM enters mainly through gauge boson self-energies, as it is
the case for extended Higgs sectors.
These constraints are implemented in \texttt{ScannerS}.
For points to be in agreement
with the experimental observation, it was required 
that the prediction of the $S$ and the $T$ parameter
are within the $2 \, \sigma$ ellipse, corresponding to
$\chi^2=5.99$ for two degrees of freedom.

\end{itemize}


\subsection{Experimental excesses}
\label{sec:excesses}

As experimental input for the signal strengths in
\citere{Biekotter:2019kde} the values
\begin{equation}
\mu_{\rm LEP} = 0.117 \pm 0.057 \quad \text{and} \quad
\mu_{\rm CMS} = 0.6   \pm 0.2   \; 
\label{mumu}
\end{equation}
were used, as quoted in \citeres{Schael:2006cr,Cao:2016uwt}
and~\cite{Sirunyan:2018aui,Shotkin:2017}. 

\smallskip
The evaluation of the signal strengths for the excesses was done in the
narrow width approximation.
For the LEP excess this is given by, 
\begin{equation}
\label{eq:mulep}
\mu_{\rm LEP} =
  \frac{\sigma_{\rm N2HDM}(e^+e^-\to Z h_1)}
       {\sigma_{\SM}(e^+e^-\to Z H)}
  \cdot
  \frac{\br_{\rm N2HDM}(h_1\to b\bar{b})}
       {\br_{\SM}(H\to b\bar{b})} =
  \left|c_{h_1 V V}\right|^2
  \frac{\br_{\rm N2HDM}(h_1\to b\bar{b})}
       {\br_{\SM}(H \to b\bar{b})} \; , 
\end{equation}
evaluated with the help of \texttt{N2HDECAY}.
For the CMS signal strength one finds, 
\begin{equation}
\label{eq:mucms}
\mu_{\rm CMS} =
  \frac{\sigma_{\rm N2HDM}(gg \to h_1)}
       {\sigma_{\SM}(gg \to H))}
  \cdot
  \frac{\br_{\rm N2HDM}(h_1 \to \gamma\gamma)}
       {\br_{\SM}(H \to \gamma\gamma)} =
  \left|c_{h_1 t \bar{t}}\right|^2
  \frac{\br_{\rm N2HDM}(h_1 \to \gamma\gamma)}
       {\br_{\SM}(H \to \gamma\gamma)} \; .
\end{equation}
The SM predictions for the branching ratios and the cross section via ggF
can be found in \citere{Heinemeyer:2013tqa}.

\begin{table}
\centering
\renewcommand{\arraystretch}{1.2}
\begin{tabular}{cccc} 
\hline
  & Decrease $c_{h_1 b \bar{b}}$ &
    No decrease $c_{h_1 t \bar{t}}$ &
    No enhancement $c_{h_1 \tau \bar{\tau}}$ \\
\hline
  type~I & \cmark $\;(\frac{R_{12}}{s_\beta})$ &
    \xmark $\;(\frac{R_{12}}{s_\beta})$ &
    \cmark  $\;(\frac{R_{12}}{s_\beta})$ \\
  type~II & \cmark  $\;(\frac{R_{11}}{c_\beta})$ &
    \cmark  $\;(\frac{R_{12}}{s_\beta})$ &
    \cmark  $\;(\frac{R_{11}}{c_\beta})$ \\
  lepton-specific & \cmark  $\;(\frac{R_{12}}{s_\beta})$ &
  \xmark  $\;(\frac{R_{12}}{s_\beta})$ &
  \xmark  $\;(\frac{R_{11}}{c_\beta})$ \\
  flipped & \cmark $\;(\frac{R_{11}}{c_\beta})$ &
  \cmark $\;(\frac{R_{12}}{s_\beta})$ &
  \xmark $\;(\frac{R_{12}}{s_\beta})$ \\
\hline
\end{tabular}
\caption{Conditions that have to be satisfied to accommodate the LEP and
CMS excesses simultaneously with a light $\cp$-even scalar $h_1$ with dominant
singlet component. In brackets we state the relevant coupling coefficients
$c_{h_1 f \bar{f}}$ for the conditions for each type.}
\label{tab:cond}
\renewcommand{\arraystretch}{1.0}
\end{table}

As can be seen from \refeqs{mumu} - (\ref{eq:mucms}), the CMS excess points
towards  the existence of a scalar with a SM-like production rate,
whereas the LEP excess demands that the scalar should have a 
squared coupling to massive vector bosons of $ \sim 0.1$ times that 
of the SM Higgs boson of the same
mass.
This suppression of the coupling coefficient $c_{h_1 V V}$
is naturally fulfilled for a singlet-like state, that acquires
its interaction to SM particles via a considerable mixing
with the SM-like Higgs boson, thus motivating the explanation
of the LEP excess with the real singlet of the N2HDM.
For the CMS excess, on the other hand, it appears to be difficult
at first sight
to accommodate the large signal strength, because one expects a
suppression of the loop-induced coupling to photons
of the same order as the one of $c_{h_1 V V}$, since in the SM the
Higgs-boson decay to photons is dominated by the $W$~boson loop.
However, it turns out that it is possible to overcompensate the
suppression of the loop-induced coupling to photons
by decreasing the
total width of the singlet-like scalar, leading to an
enhancement of the branching ratio of
the new scalar to the $\gamma \gamma$ final state.
The different types of N2HDM behave 
differently in this regard, based on how the doublet fields are
coupled to the quarks and leptons.
The general idea is summarized in \refta{tab:cond}. 

In \citere{Biekotter:2019kde} it was argued 
that only the type~II and
type~IV (flipped) N2HDM can accommodate both excesses simultaneously
using a dominantly singlet-like scalar $h_1$ at $\sim 96\gev$.
The first condition is that the coupling of $h_1$ to $b$-quarks has to be
suppressed to enhance the decay rate to $\ga\ga$,
as the total decay width at this mass
is  still dominated by the decay to $b \bar b$.
At the same time one can not decrease the coupling to $t$-quarks too much,
because the decay width to photons strongly depends on the top quark 
loop contribution (interfering constructively with the charged Higgs
contribution). 
Moreover, the ggF production cross section is dominated at leading order
by the diagram with $t$-quarks in the loop. Thus, a decreased
coupling of $h_1$ to $t$-quarks implies a lower production cross section
at the LHC.
As one can deduce from \refta{tab:cond}, in type~I and
type~III of the N2HDM, the coupling coefficients
are the same for up- and down-type quarks. Thus, it is
impossible to satisfy both of the above criteria simultaneously 
in these models. Consequently, they fail to accommodate 
both the CMS and the LEP excesses and are discarded from now on.

In \citere{Biekotter:2019kde} it is furthermore concluded that 
in type~II and IV that $|\alpha_1| \to \pi / 2$ corresponds to an
enhancement of the branching ratio to photons,
because the dominant decay width to $b$-quarks, and therefore
the total width of $h_1$, is suppressed.

A third condition, although not as significant as the other two, is
related to the coupling of $h_1$ to leptons. If it
is increased, the decay to a pair of $\tau$-leptons will be
enhanced. Similar to the decay to $b$-quarks, it will compete with
the diphoton decay and can suppress the signal strength needed for
the CMS excess. The $\tau$-Yukawa coupling is not as
large as the $b$-Yukawa coupling, so this condition is not as important
as the other two. Still, as will be reviewed below, 
it is the reason why it is easier to fit the CMS excess in the type~II model
compared to the flipped scenario.

In the scans we indicate the ``best-fit point'' referring to the
point with the smallest $\chi^2$ defined by
\begin{equation}
  \label{eq:chilepcms}
  \chi_{\rm CMS-LEP}^2 = \frac{(\mu_{\mathrm{LEP}} - 0.117)^2}{0.057^2} +
           \frac{(\mu_{\mathrm{CMS}} -   0.6)^2}{0.2^2   } \; ,
\end{equation}
quantifying the quadratic deviation w.r.t.\ the measured values, assuming
that there is no correlation between the signal strengths of
the two excesses.


\section{Results}
\label{sec:results}

In the following we will describe the analysis in the type~II (with similar
results in type~IV~\cite{Biekotter:2019kde}). 
The scalar mass eigenstate with dominant singlet-component will
be responsible for accommodating the LEP and the CMS excesses at
$\sim95$-$98\gev$. The second lightest Higgs-boson will be placed at
$\sim 125 \gev$ with the requirement that it behaves within the
uncertainties as the SM Higgs-boson.
Similar scans have been performed also for the N2HDM type~I
and~III (lepton specific), confirming that these types cannot fit well
the two excesses.

The following ranges of input parameters have been scanned:
\begin{align}
95 \gev \leq m_{h_1} \leq 98 \gev \; ,
\quad m_{h_2} = 125.09 \gev \; ,
\quad 400 \gev \leq m_{h_3} \leq 1000 \gev \; , \notag \\
400 \gev \leq m_A \leq 1000 \gev \; ,
\quad 650 \gev \leq \MHp \leq 1000 \gev \; , \notag\\
0.5 \leq \tb \leq 4 \; ,
\quad 0 \leq m_{12}^2 \leq 10^6 \gev^2 \; ,
\quad 100 \gev \leq v_S \leq 1500 \gev \; . \label{eq:ranges}
\end{align}

\begin{figure}
  \centering
  \includegraphics[width=0.8\textwidth]{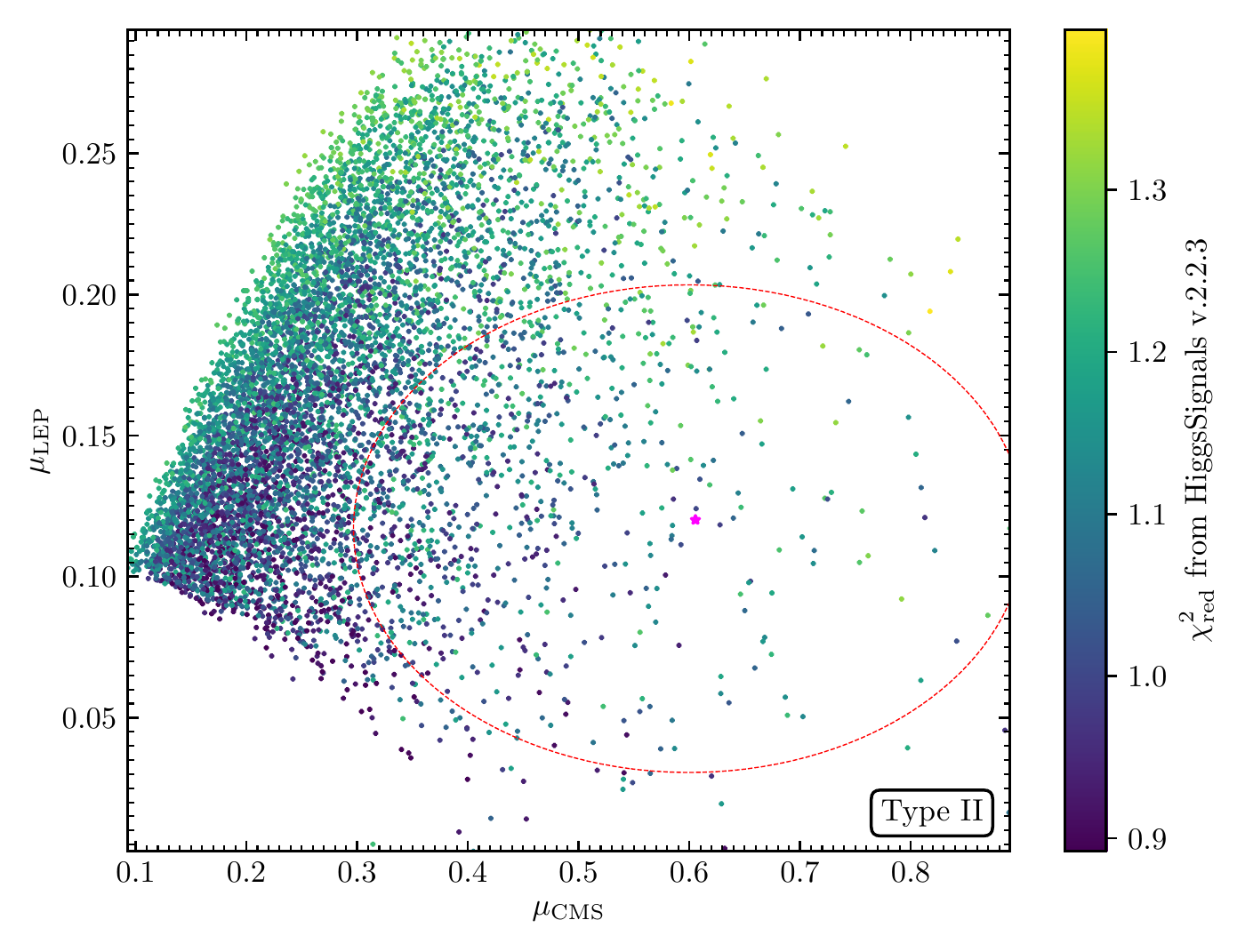}
  \caption{Type~II: the signal strengths $\mu_{\rm CMS}$ and
  $\mu_{\rm LEP}$ are shown 
  for each scan point respecting the experimental and
  theoretical constrains. The $1 \, \sigma$-region of both excesses
  is shown by the red ellipse. The colors show the
  the $\chi_{\rm red}^2$ from HiggsSignals. The best-fit point (magenta)
  has $\chi_{\rm red}^2=1.237$ with 101 observations considered.
  The lowest (highest) value of $\chi_{\rm red}^2$ inside the
  $1 \, \sigma$ ellipse is $0.9052$ ($1.3304$).}
  \label{fig:2HSplot}
\end{figure}

We show the result of the
scan in \reffi{fig:2HSplot}~\cite{Biekotter:2019kde}
in the plane of the signal strengths $\mu_{\rm LEP}$ and $\mu_{\rm CMS}$ for each
scan point, where the best-fit point w.r.t.\ the two excesses is
marked by a magenta star. It should be kept in mind that the density of
points has no physical meaning and is a pure artefact of the ``flat
prior'' in our parameter scan. 
The red dashed line corresponds to the $1\,\sig$ ellipse, i.e.,
to $\chi_{\rm CMS-LEP}^2 = 2.30$ for two degrees of freedom,
with $\chi_{\rm CMS-LEP}^2$ defined in \refeq{eq:chilepcms}.
The colors of the points indicate the reduced
$\chi^2$ from the test of the SM-like Higgs-boson properties
with \texttt{HiggsSignals}.
One sees that various points fit both excesses simultaneously while also
accommodating the properties of the SM-like Higgs boson at $125\gev$.
The lowest (hightest) value of $\MHp$ in the $1\,\sig$ ellipse is
$650.03 (964.71) \gev$, whereas the the lowest (highest) value of $\tb$
is found to be 0.797 (3.748).
It should be emphasized that the dependence of the branching ratio of
$h_1$ to diphotons, and therefore of $\mu_{\mathrm{CMS}}$,
on $\MHp$ is due to the positive correlation between $\MHp$
and the total decay width of $h_1$.
The additional contributions
to the diphoton decay width of diagrams with the charged Higgs boson
in the loop has a minor dependence on $\MHp$ for $\MHp > 650\gev$.

In \refta{tab:2best} we review the values of the free parameters
and the relevant branching ratios of the
neutral scalars
for the best-fit point of our scan,
which is highlighted with a magenta star in \reffis{fig:2HSplot}.
Remarkably, the branching ratio for the singlet-like scalar
to photons is larger than the one of the SM-like Higgs boson. As explained
in the beginning of \refse{sec:results} this is achieved by a
value of $\alpha_1 \sim \pi / 2$, 
which suppresses the decay to $b$-quarks and $\tau$-leptons, without
decreasing the coupling to $t$-quarks. Constraints from the oblique parameters
lead to a $\cp$-odd Higgs-boson mass $m_A$ or a heavy $\cp$-even
Higgs-boson mass $m_{h_3}$ close to the mass of the charged
Higgs boson.

\begin{table}
\centering
\renewcommand{\arraystretch}{1.2}
\begin{tabular}{c c c c c c c}
 $m_{h_1}$ & $m_{h_2}$ & $m_{h_3}$ & $m_A$ & $\MHp$ & & \\
 \hline
 $96.5263$ & $125.09$ & $535.86$ & $712.578$ & $737.829$ & & \\
 \hline
 \hline
 $\tb$ & $\alpha_1$ & $\alpha_2$ & $\alpha_3$ & $m_{12}^2$ & $v_S$ & \\
 \hline
 $1.26287$ & $1.26878$ & $-1.08484$ & $-1.24108$ &
   $80644.3$ & $272.72$ & \\
  \hline
  \hline
 $\br^{bb}_{h_1}$ & $\br^{gg}_{h_1}$ & $\br^{cc}_{h_1}$ &
   $\br^{\tau\tau}_{h_1}$ & $\br^{\gamma\gamma}_{h_1}$ 
 & $\br^{WW}_{h_1}$  & $\br^{ZZ}_{h_1}$ \\
 \hline
 $0.5048$ & $0.2682$ & $0.1577$ & $0.0509$ 
    & $2.582\cdot 10^{-3}$ & $0.0137$ & $1.753\cdot 10^{-3}$ \\
 \hline
 \hline
 $\br^{bb}_{h_2}$ & $\br^{gg}_{h_2}$ & $\br^{cc}_{h_2}$ &
   $\br^{\tau\tau}_{h_2}$ & $\br^{\ga\ga}_{h_2}$ &
     $\br^{WW}_{h_2}$ & $\br^{ZZ}_{h_2}$ \\
 \hline
 $0.5916$ & $0.0771$ & $0.0288$ & $0.0636$ &
   $2.153\cdot 10^{-3}$ & $0.2087$ & $0.0261$ \\
 \hline
 \hline
 $\br^{tt}_{h_3}$ & $\br^{gg}_{h_3}$ & $\br^{h_1 h_1}_{h_3}$ &
   $\br^{h_1 h_2}_{h_3}$ & $\br^{h_2 h_2}_{h_3}$ &
     $\br^{WW}_{h_3}$ & $\br^{ZZ}_{h_3}$ \\
 \hline
 $0.8788$ & $2.537\cdot 10^{-3}$ & $0.0241$ &
   $0.0510$ & $3.181\cdot 10^{-3}$ &
     $0.0261$ & $0.0125$ \\
 \hline
 \hline
 $\br^{tt}_{A}$ & $\br^{gg}_{A}$ & $\br^{Z h_1}_{A}$ &
   $\br^{Z h_3}_{A}$ & $\br^{bb}_{A}$ & & \\
 \hline
 $0.6987$ & $1.771\cdot 10^{-3}$ & $0.1008$ &
   $0.1981$ & $5.36\cdot 10^{-4}$ & $ $
\end{tabular}
\caption{Parameters of the best-fit point and branching ratios of the
lightest, second lightest and heavy $\cp$-even scalar and the $\cp$-odd
scalar in the type~II scenario.
Dimensionful parameters are given
in GeV and the angles are given in radian.}
\label{tab:2best}
\renewcommand{\arraystretch}{1.0}
\end{table}


\section{Future searches}
\label{sec:future}

\subsection{Indirect searches}
\label{sec:propsindirect}

\begin{figure}
  \centering
    \centering\includegraphics[width=0.80\textwidth]{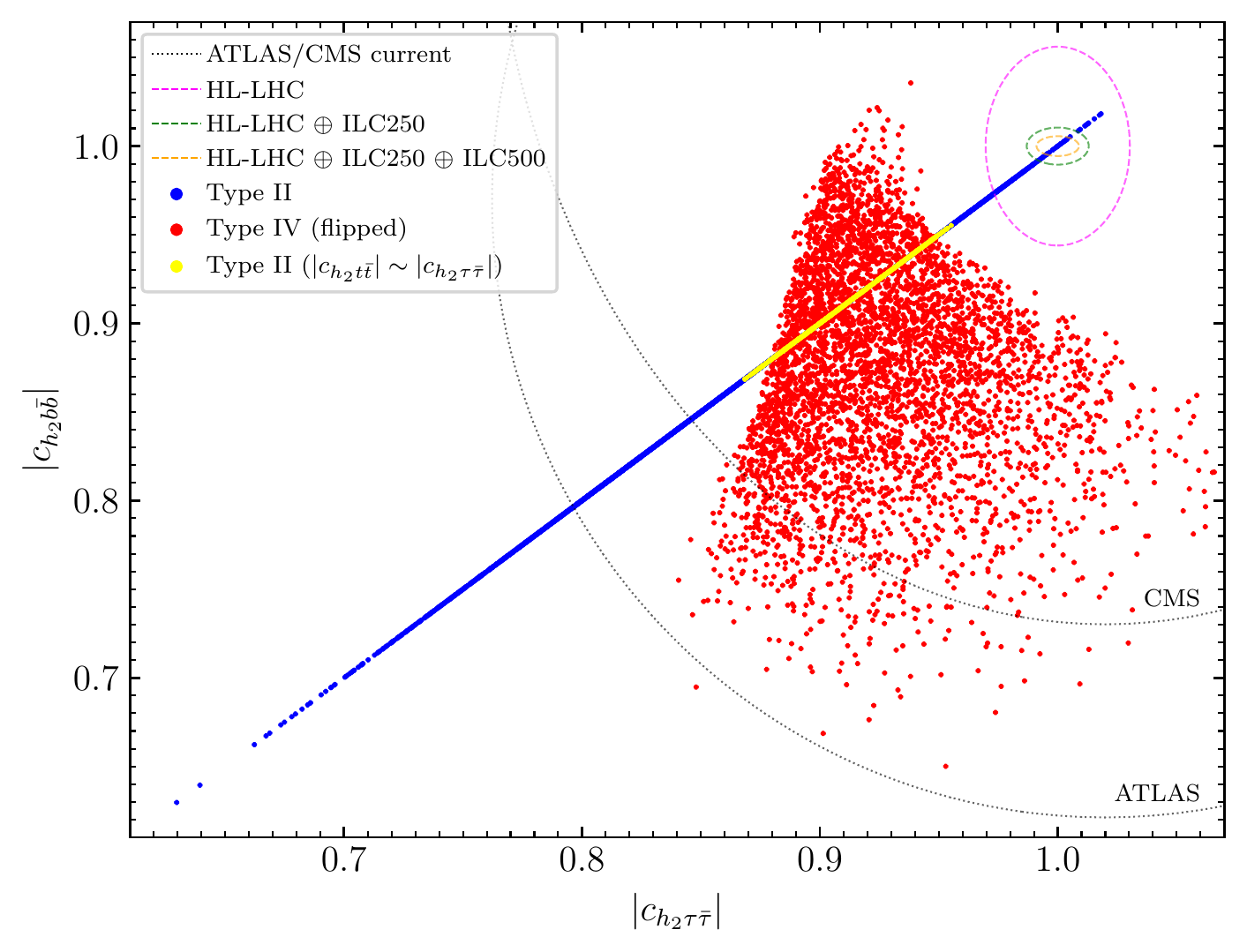}
     \centering\includegraphics[width=0.80\textwidth]{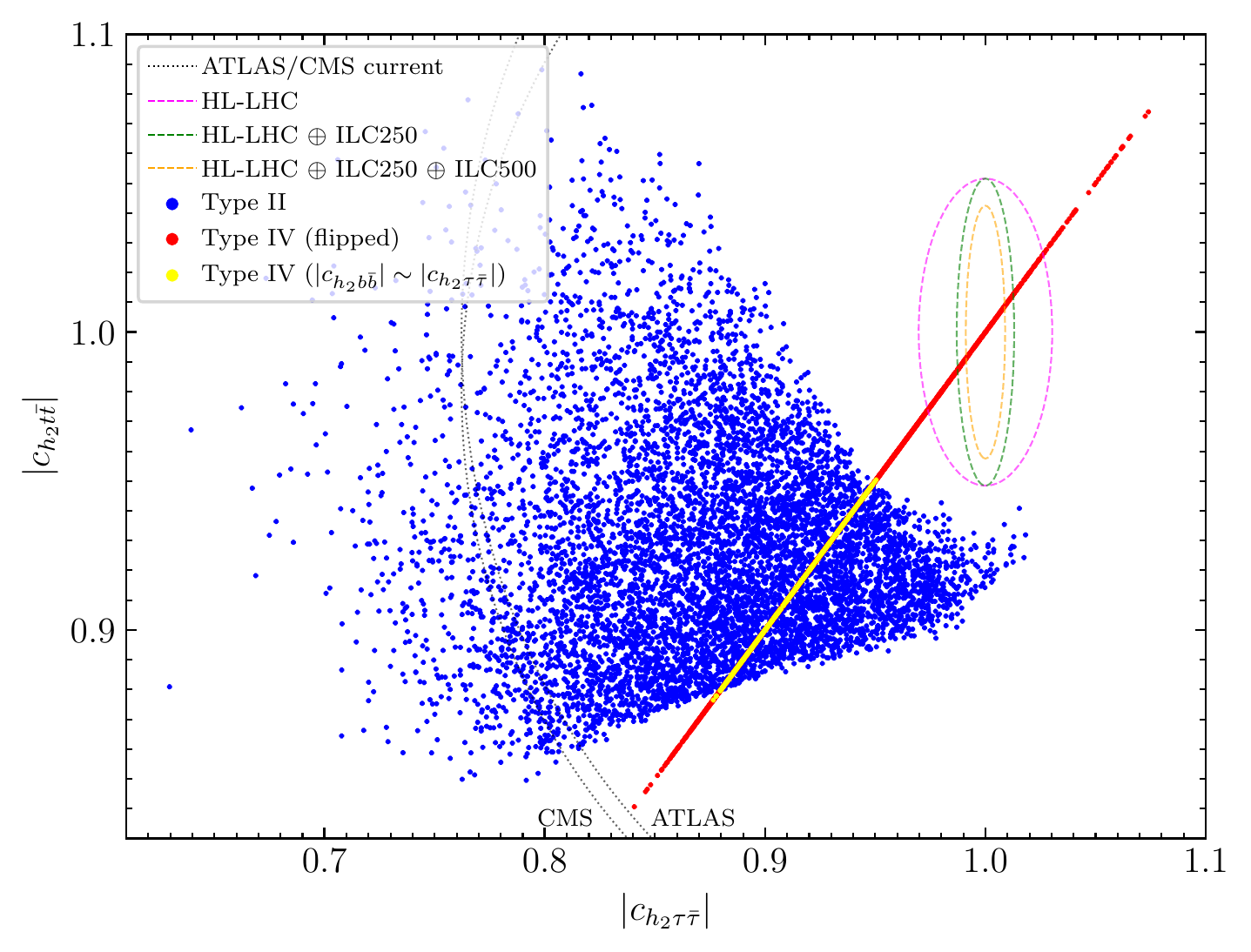}
  \caption{Scan points of the analysis in the
  type~II (\textit{blue}) and type~IV (\textit{red}) scenario in the
  $|c_{h_2 \tau \bar{\tau}}|$-$|c_{h_2 b \bar{b}}|$ plane (\textit{top}) and
  the $|c_{h_2 \tau \bar{\tau}}|$-$|c_{h_2 t \bar{t}}|$ plane (\textit{bottom}).
  In the upper plot we highlight in yellow the points of the type~II
  scenario that overlap with points from the type~IV scenario
  in the lower plot, i.e., points with
  $|c_{h_2 t \bar{t}}| \sim |c_{h_2 b \bar{b}}| \sim |c_{h_2 \tau \bar{\tau}}|$.
  In the same way in the lower plot we highlight
  in yellow the points of the type~IV
  scenario that overlap with points from the type~II scenario
  in the upper plot.
  The dashed ellipses are the projected uncertainties
  at the HL-LHC~\cite{Cepeda:2019klc} (\textit{magenta}) and the
  ILC~\cite{Bambade:2019fyw} (\textit{green} and \textit{orange})
  of the measurements of
  the coupling modifiers at the $68\%$ confidence level, assuming
  that no deviation from the SM prediction will be found
  (more details in the text).
  We also show with the dottet black lines the $1 \, \sigma$ ellipses
  of the current measurements from
  CMS~\cite{Sirunyan:2018koj} and
  ATLAS~\cite{ATLAS-CONF-2018-031}.}
  \label{fig:cplprosp}
\end{figure}

Currently, uncertainties on the measurement of the coupling
strengths of the SM-like Higgs boson at the LHC are still large,
i.e., at the $1 \, \sigma$-level they are of the same order as the
modifications of the couplings present
in our analysis in the
N2HDM~\cite{Khachatryan:2016vau,ATLAS-CONF-2018-031,Sirunyan:2018koj}.
In the future, once the complete $300\, \ifb$ collected at the LHC are analyzed,
the constraints on the couplings of the SM-like
Higgs boson will benefit from the reduction of statistical uncertainties.
Even tighter constraints are expected from the LHC after the
high-luminosity upgrade (HL-LHC), when the planned amount
of $3000\, \ifb$ integrated luminosity will have been
collected~\cite{Dawson:2013bba}. 
Finally, a future linear $e^+ e^-$ collider like the ILC, CLIC, FCC-ee
or CepC could
improve the precision measurements of the {Higgs-boson couplings
even further~\cite{Dawson:2013bba,Drechsel:2018mgd}, where we will use
ILC numbers for illustration.
At an $e^+e^-$ collider the cross section of the 
Higgs boson can be measured independently,
and the total width (and therefore also the coupling
modifiers) can be reconstructed without model assumptions.

Several studies have been performed to estimate the future constraints
on the coupling modifiers of the SM-like Higgs boson at the
LHC~\cite{Dawson:2013bba,CMS:2013xfa,Tricomi:2015nrd,ATL-PHYS-PUB-2014-016,Slawinska:2016zeh} and the
ILC~\cite{Dawson:2013bba,Asner:2013psa,Ono:2013sea,Durig:2014lfa,Fujii:2017vwa,Bambade:2019fyw,Cepeda:2019klc},
assuming that no deviations from the SM predictions will be found.
Here, we review the comparison of the scan points to the expected precisions of
the HL-LHC and the ILC as they are reported
in \citeres{Bambade:2019fyw,Cepeda:2019klc},
neglecting possible correlations of the coupling modifiers.
The results are shown in \reffi{fig:cplprosp}~\cite{Biekotter:2019kde}.

We plot the effective
coupling coefficient of the SM-like Higgs boson $h_2$ to $\tau$-leptons
on the horizontal axis against the coupling coefficient to
$b$-quarks (top) and to $t$-quarks (bottom) for both types.
These points passed all the experimental and theoretical constraints, including
the verification of SM-like Higgs-boson properties in agreement
with LHC results using \texttt{HiggsSignals}.
In the top plot the blue points lie on a diagonal line, because in
type~II the coupling to leptons and to down-type quarks scale identically,
while in the bottom plot the red points representing
the type~IV scenario lie on the diagonal,
because there the lepton-coupling scales in the same
way as the coupling to up-type quarks.

In \reffi{fig:cplprosp} the current measurements
on the coupling modifiers by ATLAS~\cite{ATLAS-CONF-2018-031}
and CMS~\cite{Sirunyan:2018koj}
are shown as black ellipses.
The magenta ellipse in each plot shows the expected precision of the measurement
of the coupling coefficients at the $1 \, \sigma$-level at the
HL-LHC from \citere{Cepeda:2019klc}.
The current uncertainties and the HL-LHC analysis are based on the
coupling modifier, or $\kappa$-framework, in which the tree-level
couplings of the SM-like Higgs boson to
vector bosons, the top quark, the bottom quark, the $\tau$ and the $\mu$ lepton,
and the three loop-induced
couplings to $\ga\ga$, $gg$ and $Z\gamma$ receive a factor $\kappa_i$
quantifying 
potential modifications from the SM predictions. These modifiers are then
constrained using a global fit to projected HL-LHC data assuming no
deviation from the SM prediction will be found. The uncertainties found
for the $\kappa_i$ can directly be applied to the future precision
of the coupling modifiers $c_{h_i \dots}$ we use in our paper.
We use the uncertainties given under the assumptions that
no decay of the SM-like Higgs boson to BSM particles is present,
and that current systematic uncertainties will be reduced in addition
to the reduction of statistical uncertainties due to the increased statistics.

The green and the orange ellipses show the corresponding expected
uncertainties when the HL-LHC results are combined with projected
data from the ILC after the $250\gev$ phase and
the $500\gev$ phase, respectively, taken from \citere{Bambade:2019fyw}.
Their analysis is based on a pure effective field theory calculation,
supplemented by further assumptions to facilitate the combination with
the HL-LHC projections in the $\kappa$-framework. In particular, in
the effective field theory approach the vector boson couplings can
be modified beyond a simple rescaling. This possibility was excluded
by recasting the fit setting two parameters related to the couplings
to the $Z$-boson and the $W$-boson to zero
(for details we refer to \citere{Bambade:2019fyw}).

Remarkably, the expected constraints from the HL-LHC and the ILC will
strongly reduce the allowed parameter spaces and allow a clear test of
the models under consideration. 
Independent of the type of the N2HDM, we can see comparing both
plots in \reffi{fig:cplprosp}, that there is not a single scan
point that coincides with the SM prediction regarding the three
coupling coefficients shown. This implies that, once these
couplings are measured precisely by the HL-LHC and the ILC,
a deviation of the SM prediction has to be measured in at least one
of the couplings, if our explanation of the excesses is correct.
Accordingly, if no deviation from the SM prediction regarding these
couplings will be measured, our explanation would be ruled out entirely.

Furthermore, in case a deviation from the SM prediction will be found,
the predicted scaling behavior of the coupling coefficients in the type
II scenario (upper plot) and the type~IV scenario (lower plot), might
lead to distinct possibilities for the two models to accommodate these
possible deviations. In this case, precision measurements of the
SM-like Higgs boson couplings could be used to exclude one of the
two scenarios.
This is true for all points except the ones highlighted
in yellow in \reffi{fig:cplprosp}. The yellow points are a subset of points
of our scans that, if such deviations of the SM-like Higgs boson
couplings will be measured, could correspond to a benchmark point
of both the scan in the type~II and the type~IV scenario.
However, note that this subset of points is confined to the diagonal
lines of both plots, and thus corresponds to a very specific subset
of the overall allowed parameter space.
For the type~II scenario, in the upper plot,
the yellow points are determined by the additional
constraint that $|c_{h_2 t \bar{t}}| \sim |c_{h_2 \tau \bar{\tau}}|$,
which is exactly true in the type~IV scenario.
For the type~IV scenario, in the lower plot,
the yellow points are determined by the additional
constraint that $|c_{h_2 b \bar{b}}| \sim |c_{h_2 \tau \bar{\tau}}|$,
which is exactly true in the type~II scenario.

\begin{figure}
  \centering
    \centering\includegraphics[width=\textwidth]{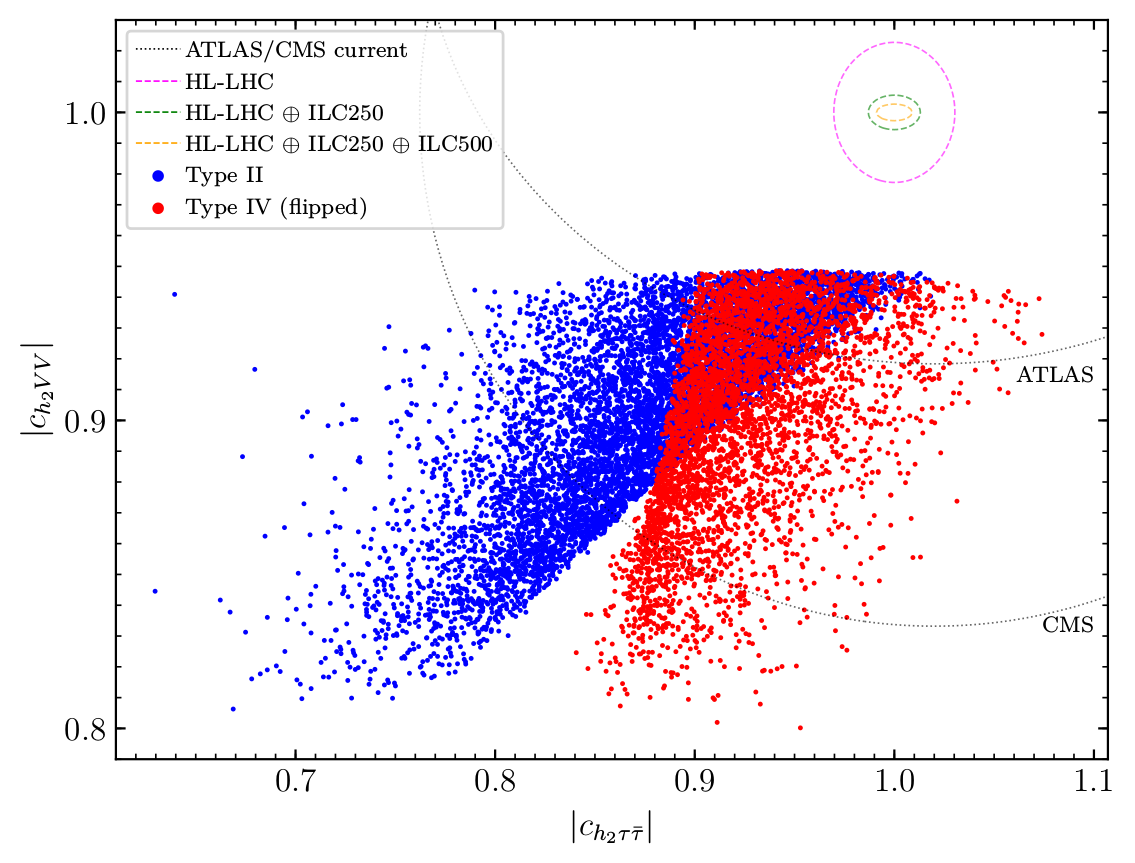}
  \caption{As in Fig.~2 but with $|c_{h_2 V V}|$
  on the vertical axis.}
  \label{fig:cplprospvv}
\end{figure}

For completeness we show in \reffi{fig:cplprospvv} the
absolute value of the coupling
modifier of the SM-like Higgs boson w.r.t.\ the vector boson couplings
$|c_{h_2 V V}|$ on the vertical axis. Again, the parameter points of both
types show deviations larger than the projected experimental uncertainty
at HL-LHC and ILC.
The deviations in $|c_{h_2VV}|$ are even stronger than for the
couplings to fermions. A $2\,\sig$ deviation from the SM prediction is expected
with HL-LHC accuracy. At the ILC a deviation fo more than $5\,\sig$
would be visible.
As mentioned already, a suppression of the
coupling to vector bosons is explicitly
expected by demanding $\Sigma_{h_2} \geq 10\%$. However,
since points with lower singlet component cannot accommodate
both excesses, this does not contradict the conclusion that
the explanation of both excesses can be
probed with high significance with future Higgs-boson coupling
measurements.


\subsection{Direct searches}
\label{sec:propsdirect}

To start with, the diphoton bump which has persisted through LHC Run\,I and II
is worth exploring in additional Higgs boson searches of future runs of the LHC.
Furthermore, the search for charged Higgs bosons appears promising in the
region of low $\tb$. 
Searches at the \mbox{(HL-)LHC} will yield strong constraints or (hopefully)
discover signs of a charged Higgs-boson in the region between $600 \gev$ and
$950 \gev$. Prospects for a $5\,\sig$ discovery
in the charged Higgs-boson searches in the $tb$ decay mode can be
found in \citere{Guchait:2018nkp}.

\begin{figure}
  \centering
  \includegraphics[width=0.8\textwidth]{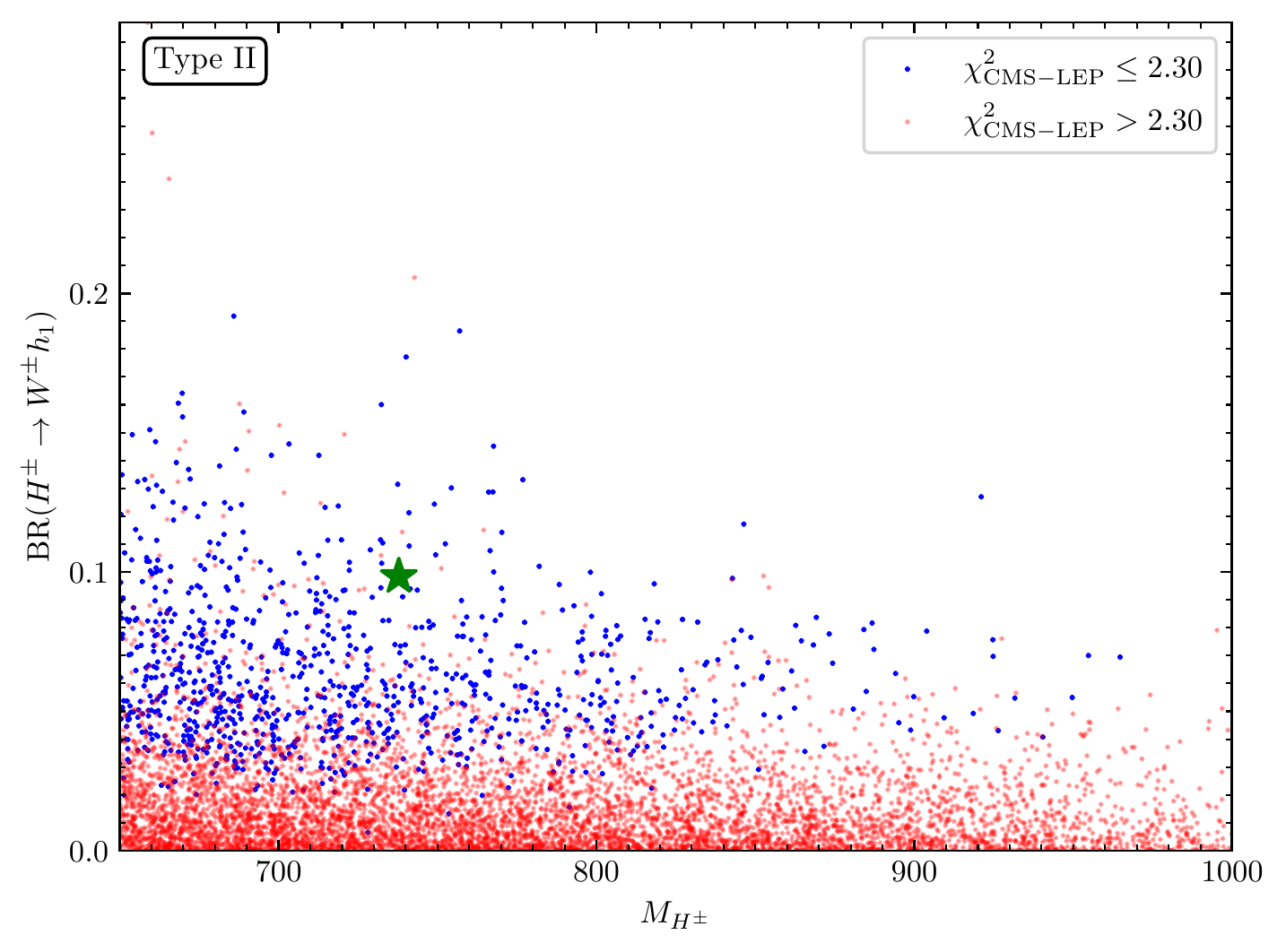}
  \caption{Type~II: The branching ratios
  $\br(\Hpm \rightarrow W^\pm h_1)$
  are shown for each parameter point inside (\textit{blue}) and outside
  (\textit{red}) the $1\,\sigma$ ellipse regarding the CMS and the LEP excesses.
  The best-fit point is marked by the green star.}
  \label{fig:2BrHpm1}
\end{figure}
\begin{figure}
  \centering
  \includegraphics[width=0.8\textwidth]{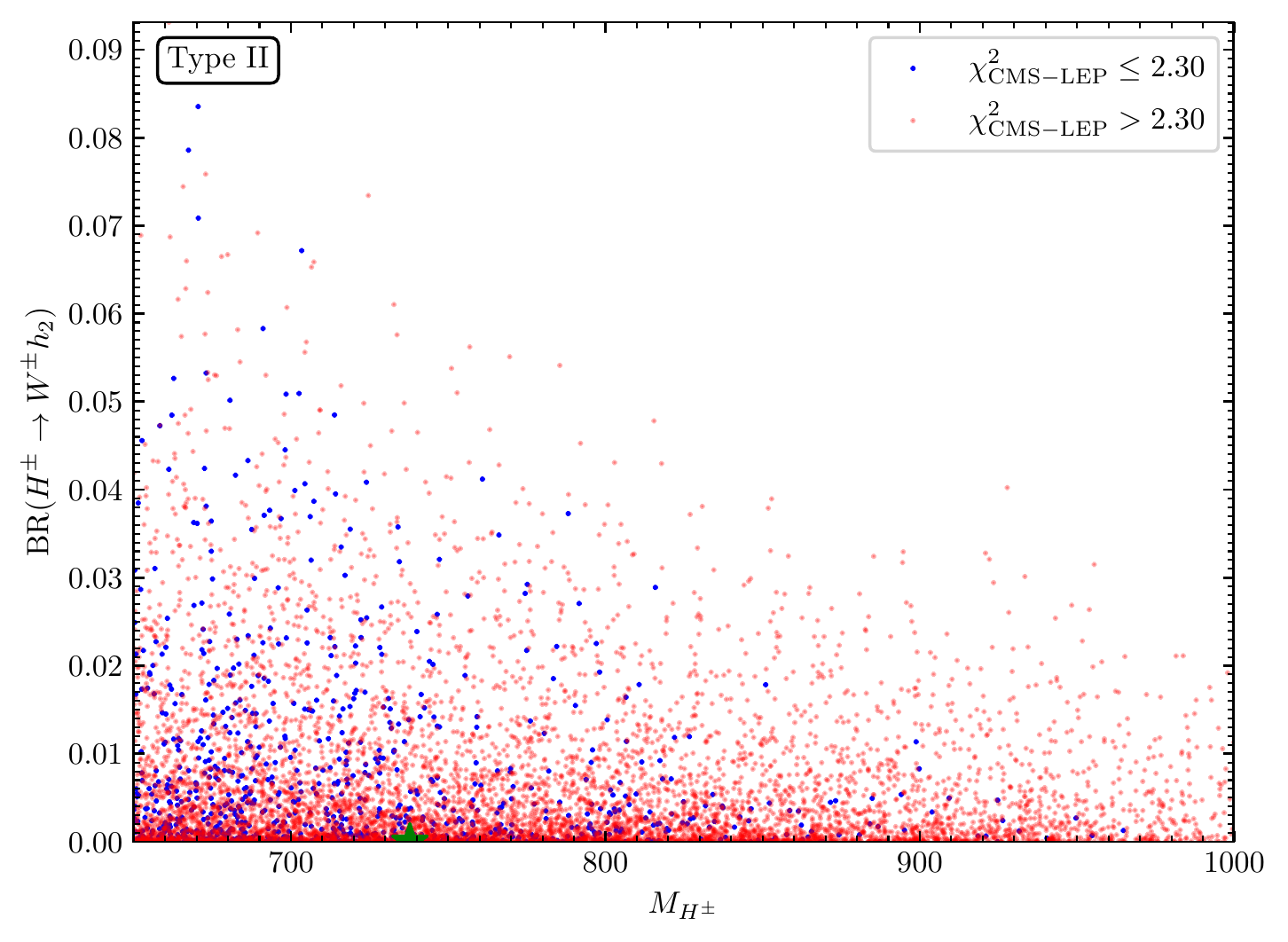}
  \caption{Type~II: Same as in Fig.~4 for
  $\br(\Hpm \rightarrow W^\pm h_2)$.}
  \label{fig:2BrHpm2}
\end{figure}
\begin{figure}
  \centering
  \includegraphics[width=0.8\textwidth]{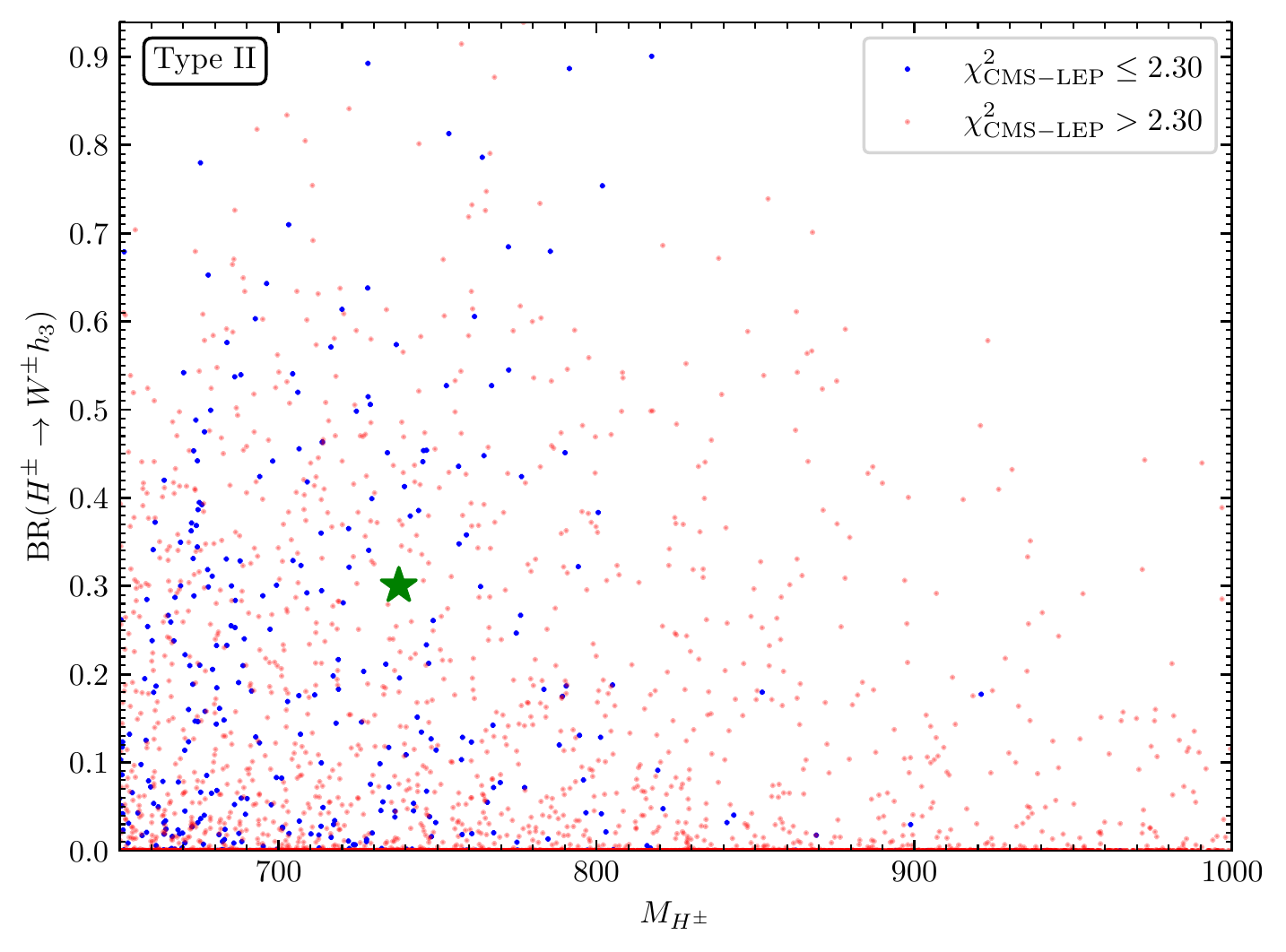}
  \caption{Type~II: Same as in Fig.~4 for
  $\br(\Hpm \rightarrow W^\pm h_3)$.}
  \label{fig:2BrHpm3}
\end{figure}

Since the charged Higgs boson is rather heavy due to
the constraints from flavor physics, exotic signals
at colliders can be expected from the decay of the charged
Higgs boson into a $W$ boson and a neutral Higgs bosons.
We show the corresponding branching ratios in \reffi{fig:2BrHpm1},
\ref{fig:2BrHpm2} and \ref{fig:2BrHpm3} for the decays
of $\Hpm$ into $W^\pm$ and $h_1$, $h_2$ and $h_3$, respectively.
The blue points are the ones that lie inside the $1\,\sigma$ ellipse
of $\mu_{\mathrm{LEP}}$ and $\mu_{\mathrm{CMS}}$.
The decays into the two light Higgs bosons is always kinematically
allowed. However, as one can see in \reffi{fig:2BrHpm3},
if the decay to the heavy Higgs boson $h_3$ opens up kinematically, it
is usually the dominant of the three, and competes with ordinary decay modes
of $\Hpm$ into a pair of $tb$ quarks. The smallest branching ratio
for the mass range of $\MHp$ in our scan is the one
to the SM-like Higgs boson $h_2$, which is minimized
in the limit of $h_2$ becoming SM-like.
Concerning the decay to the lightest Higgs boson $h_1$, a correlation
is visible. The points explaining both excesses within the $1\,\sigma$
uncertainty have larger branching fractions. In order for this
decay to happen, $h_1$ needs a sizable doublet component, otherwise
it would not couple to the $W$ boson. The doublet component is,
as explained before, also necessary for $h_1$ to contribute to
the signal strengths at LEP and CMS.

The prospects for the searches for the heavy neutral Higgs bosons,
decaying dominantly to $t \bar t$, may also be promising. However, we
are not aware of corresponding HL-LHC projections.

$e^+ e^-$ colliders, on the other hand 
show good prospects for the search of light
scalars \cite{Wang:2018fcw,Drechsel:2018mgd}. 
The main production channel in the mass and energy range 
that we are interested in is the Higgs-strahlung process
$e^+ e^- \to \phi Z$, where $\phi$ is the
scalar being searched for.
The LEP collaboration has previously performed such
searches~\cite{Barate:2003sz},
which resulted in the $2\,\sig$ excess given by $\mu_{\rm LEP}$.
These searches were limited by the low luminosity of LEP.
However, the ILC, with its much higher luminosity and the possibility 
of using polarized beams, has a substantially higher 
potential to discover the light scalars.
The searches performed at LEP can be divided into two categories:
the 'traditional method', where studies are based on the decay mode
$\phi \to b \bar b$ along with $Z$ decays to $\mu^+ \mu^-$
final states. 
This method introduces certain amount of model dependence 
into the analysis because of the reference to a specific decay mode of $\phi$. 
The more model independent 'recoil technique' used by 
the OPAL collaboration of LEP looked for light states 
by analyzing the 
recoil mass distribution of the di-muon system 
produced in $Z$ decay \cite{Abbiendi:2002qp}. 

In \reffi{fig:2ilc}~\cite{Biekotter:2019kde} the 
bounds from the LEP as well as the projected 
bounds from the ILC searches for light scalars in type~II 
N2HDM scenarios are shown. The lines indicating the ILC reach for
a $\sqrt s = 250 \gev$ machine with beam polarizations 
$(P_{e^-}, P_{e^+})$ of $(-80\%,+30\%)$ and an integrated
luminosity of 2000\,$\ifb$ are as evaluated 
in \citere{Drechsel:2018mgd}. The quantity $S_{95}$ used in their
analysis corresponds to an upper limit at the $95\%$ confidence level 
on the cross section times branching ratio generated within the
'background only' hypothesis, where the cross section has been
normalized to the reference SM-Higgs 
cross section and the BRs have been assumed to be as in the SM
(with a Higgs boson of the same mass). Consequently, we take the
obtained limits to be valid for the total cross section times branching ratio.
The colored points shown in \reffi{fig:2ilc} 
are the points of the scans in the type~II scenario
satisfying all the theoretical and experimental constraints.
The plot show that the parameter points 
of the scans can completely be covered by searches
at the ILC for additional Higgs-like scalars.
Depending on $c_{h_1 V V}$, i.e., the light Higgs-boson production
cross section, the $h_1$ can be produced and analyzed in detail
at the ILC. 

\begin{figure}
  \centering
    \centering\includegraphics[width=\textwidth]{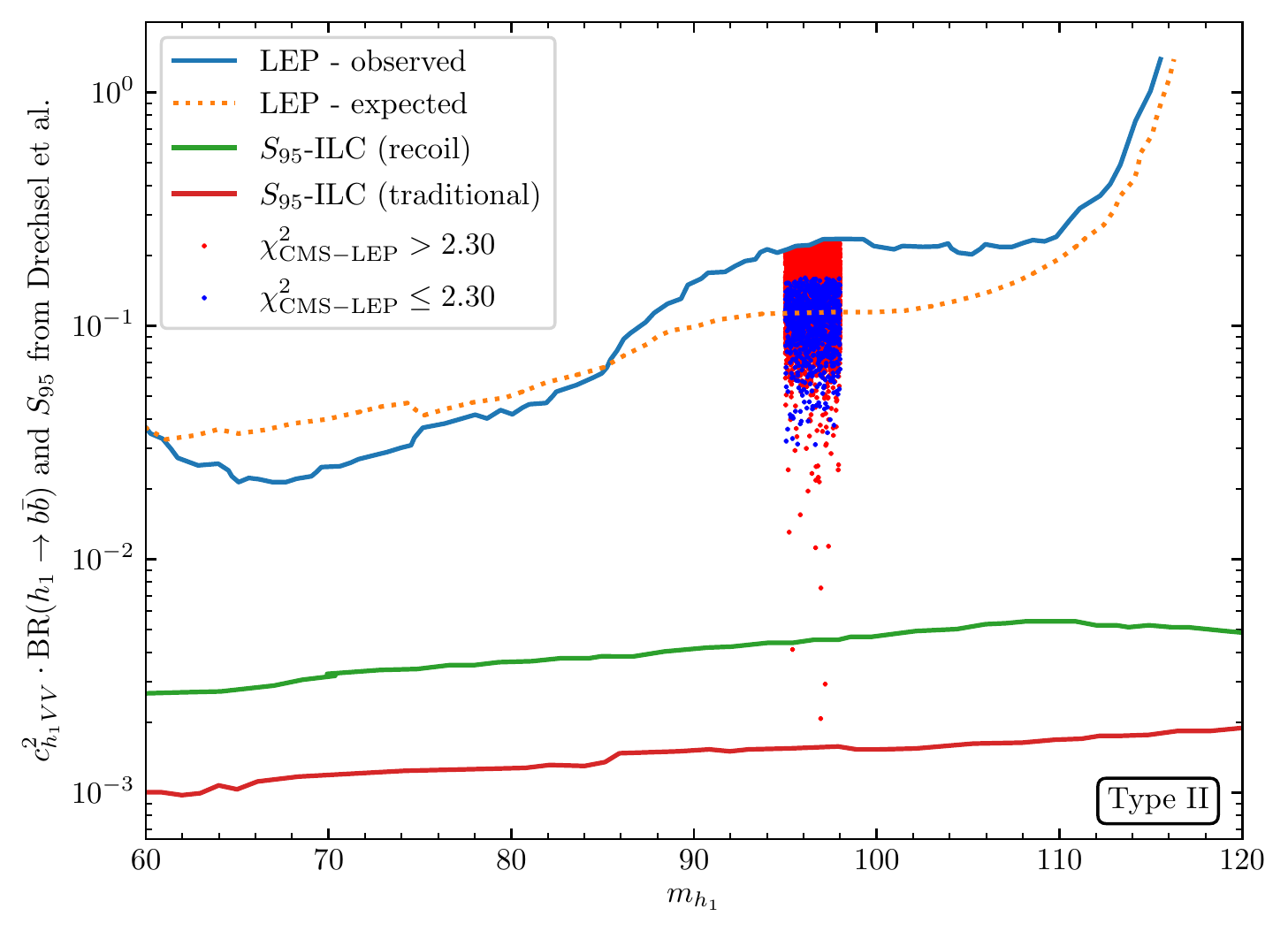}
  \caption{The $95\%$ CL expected (\textit{orange dashed}) and
  observed (\textit{blue}) upper bounds on the Higgsstrahlung production
  process with associated decay of the scalar to a pair of
  bottom quarks at LEP~\cite{Barate:2003sz}.
  Expected $95\%$ CL upper limits on the Higgsstrahlung production
  process normalized to the SM prediction $S_{95}$ at
  the ILC using the traditional (\textit{red}) and the
  recoil technique (\textit{green}) as described in the
  text~\cite{Drechsel:2018mgd}. We also show the points of our scan
  in the type~II scenario which lie within (\textit{blue}) and
  outside (\textit{red}) the $1 \, \sigma$ ellipse
  regarding the CMS and the LEP
  excesses.}
  \label{fig:2ilc}
\end{figure}


\section{Supersymmetric realizations}
\label{sec:susy}

\sloppy

In \refse{sec:excesses} it was demonstrated that due to the structure of
the couplings of the Higgs 
doublets to fermions only two types of the N2HDM, type~II and type~IV
(flipped), can fit simultaneously the two excesses. 
Due to the different coupling to leptons in type~II and type~IV, in
general larger values of $\mu_{\rm CMS}$ can be reached in the former,
and the CMS excess can be fitted ``more naturally'' in the type~II N2HDM. 
Incidentally, this is exactly the Higgs sector that is required by
supersymmetric models. On the other hand, 
in \citere{Bechtle:2016kui} it was shown that the MSSM cannot
explain the CMS excess in the diphoton final state. This can be traced
back to the ``too rigid'' structure of the 2HDM (type~II) strucure of the
Higgs-boson sector in the MSSM. SUSY models that can potentially explain
both excesses simultaneously, consequently, should contain (at least) an
additional Higgs singlet.

Going beyond the MSSM, a well-motivated extension is given by
the Next-to-MSSM (NMSSM), see
~\cite{Ellwanger:2009dp,Maniatis:2009re} for reviews. 
In the NMSSM a new singlet superfield is introduced, which only couples to the
Higgs- and sfermion-sectors, giving rise to an effective $\mu$-term. 
In the $\cp$-conserving case the NMSSM Higgs sector consists of
three $\cp$-even Higgs bosons,
$h_i$ ($i = 1,2,3$), two $\cp$-odd Higgs bosons, $a_j$ ($j = 1,2$),
and the charged Higgs boson pair $H^\pm$. 
In the NMSSM not only the
lightest but also the second lightest $\cp$-even Higgs boson
can be interpreted as the signal observed at about $125~\gev$, see,
e.g., \cite{King:2012is,Domingo:2015eea}.
In \citere{Domingo:2018uim} it was demonstrated that the NMSSM can indeed
simultaeneously satisfy the two excesses mentioned above.
In this case, the Higgs boson at $\sim 96 \gev$ has a large 
singlet component, but also a sufficiently large doublet 
component to give rise to the two excesses. 

A natural extension of the NMSSM is the \mnSSM, in which the 
singlet superfield is interpreted as a right-handed neutrino 
superfield~\cite{LopezFogliani:2005yw,Escudero:2008jg},
see \citeres{Munoz:2009an,Munoz:2016vaa,Ghosh:2017yeh} for reviews.
The \mnSSM\ is the simplest extension of the MSSM that can provide 
massive neutrinos through a see-saw mechanism at the electroweak scale.
A Yukawa coupling for right-handed neutrinos of the order of the
electron Yukawa coupling is introduced that induces the explicit
breaking of $R$-parity. 
Also in the \mnSSM\ the signal at $\sim 125 \gev$ can
be interpreted as the lightest or the second lightest $\cp$-even scalar.
In \citere{Biekotter:2017xmf} the ``one generation case'' (only one
generation of massive neutrinos) was analyzed. In this case 
Higgs-boson sector of the \mnSSM\ effectively resembles
the Higgs-boson sector in the NMSSM. 
In \citere{Biekotter:2017xmf} it was found that also the \mnSSM\
can fit the CMS and the LEP excesses simultaneously. In this case the
scalar at $\sim 96 \gev$ has a large right-handed sneutrino component.
The three generation case (i.e.\ with three generations of massive
neutrinos) is currently under investigation~\cite{munuSSM-3g}.


\section {Conclusions}
\label{sec:conclusion}

A $\sim 3\,\sigma$ excess (local)
in the diphoton decay mode at $\sim 96 \gev$ was reported by
CMS, as well as a $\sim 2\,\sigma$ excess (local) in the
$b \bar b$ final state at LEP in the same mass range.
We reviewed the interpretation this possible signal as a Higgs boson in the
2~Higgs Doublet Model with an additional real Higgs singlet
(N2HDM)~\cite{Biekotter:2019kde}. 

All relevant constraints were included in the analysis. These are
theoretical constraints from perturbativity and the requirement that the
minimum of the Higgs potential is a global minimum. We take into account
the direct searches for additional Higgs bosons from LEP. the Tevatron
and the LHC, as well as the measurements of the properties of the Higgs
boson at $\sim 125 \gev$. We furthermore include bounds from flavor
physics and from electroweak precision data. 

It was demonstrated that due to the structure of the couplings of the Higgs
doublets to fermions only two types of the N2HDM, type~II and type~IV
(flipped), can fit simultaneously the two excesses. On the other hand,
the other two types, type~I and type~III (lepton specific), cannot be
brought in agreement with the two excesses. 
Subsequently, the free parameters in the two favored versions
of the N2HDM were scanned, where the results are similar in both
scenarios. It was found 
that the lowest possible values of $\MHp$ above $\sim 650 \gev$ and
$\tb$ just above~1 are favored. The reduced $\chi^2$ from the
Higgs-boson measurements is found roughly in the range 
$0.9 \lesssim  \chi_{\rm red}^2 \lesssim 1.3$. 
Due to the different coupling to leptons in type~II and type~IV, in
general larger values of $\mu_{\rm CMS}$ can be reached in the former,
and the CMS excess can be fitted ``more naturally'' in the type~II N2HDM. 
Incidentally, this is exactly the Higgs sector that is required by
supersymmetric models. 

It was analyzed how the favored scenarios can be tested at future
colliders. The (HL-)LHC will continue the searches/measurements in the
diphoton final state. But apart from that we are not aware of other
channels for the light Higgs boson that could be accessible. 
Concerning the searches for heavy N2HDM Higgs bosons,
particularly interesting are the prospects for charged Higgs bosons.
For the low $\tan\beta$ values favored in our analysis, these
searches have the best potential to discover a new heavy Higgs boson
at the LHC Run III or the HL-LHC.
The prospects for the searches for the heavy neutral Higgs bosons,
decaying dominantly to $t \bar t$, may also be promising. However, we
are not aware of corresponding HL-LHC projections.

A future $e^+e^-$ collider, such as the ILC, CLIC, FCC-ee or CepC, will
be able to produce the light Higgs state at $\sim 96\gev$
in large numbers and consequently study its decay
patterns. Similarly, it was demonstrated that
the high anticipated precision in the
coupling measurements of the $125\gev$ Higgs boson
at the ILC, CLIC, FCC-ee, or CepC will allow to
find deviations w.r.t.\ the SM values if the N2HDM with a $\sim 96 \gev$
Higgs boson is realized in nature.
Here the coupling of the SM-like Higgs boson to the massive SM gauge bosons
appears to be particularly promising.

Based on the fact that type~II can fit the two excesses ``most
naturally'', we reviewed briefly two SUSY solutions to the two excesses:
these are models with two Higgs doublets and (effectively) one Higgs
singlet: the NMSSM and the  (one-generation case) \mnSSM. In both
models, despite the additional SUSY constraints on the Higgs-boson
sector, the two excesses can indeed be fitted simultaneously.


\subsection*{Acknowledgements}

S.H.\ thanks the organizers of the Corfu Summer Institute 2018 ``School
and Workshops on Elementary Particle Physics and Gravity'' (CORFU2018)
for the warm hospitality, the inspiring atmosphere and the excellent
``local specialities''.\\
We thank 
R.~Santos, 
T.~Stefaniak
and
G.~Weiglein
for helpful discussions. M.C.\ thanks D.~Azevedo for discussions regarding
\texttt{ScannerS}.
The work  was supported in part by the MEINCOP (Spain) under 
contract FPA2016-78022-P and in part by the AEI
through the grant IFT Centro de Excelencia Severo Ochoa SEV-2016-0597. 
The work of T.B.\ and S.H.\ was 
supported in part by the Spanish Agencia Estatal de
Investigaci\'on (AEI), in part by
the EU Fondo Europeo de Desarrollo Regional (FEDER) through the project
FPA2016-78645-P, in part by the ``Spanish Red Consolider MultiDark''
FPA2017-90566-REDC.
The work of T.B.\ was
funded by Fundaci\'on La Caixa under `La Caixa-Severo Ochoa' international
predoctoral grant.




\begingroup\raggedright\endgroup


\end{document}

%% file: paperdef.tex
\newcommand\tb{\tan\beta}

\newcommand\ReDiag{\mathop{%
  \raise .5pt\hbox{[}%
  \widetilde{\mathrm{Re}}%
  \raise .5pt\hbox{]}}}
\newcommand\ReOffDiag{\mathop{%
  \raise .5pt\hbox{$\llbracket$}%
  \widetilde{\mathrm{Re}}%
  \raise .5pt\hbox{$\rrbracket$}}}

\newcommand\SW{s_\mathrm{w}}
\newcommand\CW{c_\mathrm{w}}
\newcommand\MW{M_W}
\newcommand\MZ{M_Z}

\newcommand\MHp{M_{H^\pm}}

\newcommand\refeq[1]{Eq.~(\ref{#1})}
\newcommand\refeqs[1]{Eqs.~(\ref{#1})}
\newcommand\refta[1]{Tab.~\ref{#1}}
\newcommand\refse[1]{Sect.~\ref{#1}}

\newcommand\citere[1]{Ref.~\cite{#1}}
\newcommand\citeres[1]{Refs.~\cite{#1}}

\newcommand{\SM}{\mathrm{SM}}

\newcommand{\mnSSM}{\ensuremath{\mu\nu\mathrm{SSM}}}

\newcommand{\CP}{{\cal CP}}
\newcommand{\cp}{{\CP}}

\newcommand{\tev}{\,\, \mathrm{TeV}}
\newcommand{\gev}{\,\, \mathrm{GeV}}

\newcommand{\Hpm}{H^\pm}

\newcommand\fb{\ensuremath{\mbox{fb}}}

\newcommand\ifb{\ensuremath{\fb^{-1}}}

\newcommand{\br}{\text{BR}}

\newcommand{\sig}{\sigma}

\def\reffi#1{\mbox{Fig.~\ref{#1}}}
\def\reffis#1{\mbox{Figs.~\ref{#1}}}

\def\ga{\gamma}

\definecolor{Orange}{rgb}{0.8,0.5,0.}
\definecolor{Purple}{rgb}{0.5,0.,0.5}
\definecolor{Lightblue}{cmyk}{0.9,0.1,0.1,0.3}
\definecolor{dgelborange}{cmyk}{0.,0.3,0.5, 0.}
\definecolor{Lila}{rgb}{0.5,0.,1}